\documentclass[aps,pre,twocolumn,groupedaddress,longbibliography]{revtex4-2}

\usepackage{graphicx}
\usepackage{natbib}
\usepackage{subfigure}

\usepackage{color}
\usepackage{multirow}
\usepackage{amssymb}
\usepackage{booktabs}
\usepackage{algorithm}
\usepackage{todonotes}
\usepackage{lineno}
\usepackage[final]{hyperref} 
\usepackage{amsmath}
\usepackage{cleveref}

\usepackage[inline]{trackchanges} 
\addeditor{LK}
\addeditor{JA}
\addeditor{BG}
\addeditor{LC}

\begin{document}

\title{On the fractal dimension of non-Newtonian Hele-Shaw flow subject to Saffman-Taylor instability}

\author{J. Adriazola}
\affiliation{Department of Mathematics, Southern Methodist University, Dallas, Texas, 75205}

\author{B. Gu}
\affiliation{Department of Mathematical Sciences, Worcester Polytechnic Institute, Worcester, Massachusetts, 01609}


\author{L.J. Cummings and L. Kondic}
\email[]{kondic@njit.edu}
\homepage[]{https://cfsm.njit.edu}
\affiliation{Department of Mathematical Sciences, New Jersey Institute of Technology, Newark, New Jersey 07102, USA}

\date{\today}

\begin{abstract}
We introduce a discrete numerical method based on the diffusion-limited aggregation (DLA) approach to simulate two-fluid Hele-Shaw flow subject to the Saffman-Taylor interfacial instability, in the case where the displaced fluid is non-Newtonian. Focusing on fluids for which the most relevant non-Newtonian aspect of the thin-gap flow is shear-thinning, we introduce a history-dependent aspect into the algorithm, modeling shear-rate-dependent fluid viscosity. The main finding is that the morphology of the emerging patterns, characterized by the fractal dimension, is modified in a nontrivial manner by the shear-thinning nature of the displaced fluid. In particular, we consistently find that shear-thinning leads to the formation of patterns characterized by a smaller fractal dimension, compared to the corresponding Newtonian fluid.
\end{abstract}

\maketitle
\section{Introduction}
\label{sec:intro}

\begin{figure}[h!]
  \centering
  \subfiguretopcaptrue
  \subfigure[\qquad\qquad\qquad\qquad\qquad\qquad]{\includegraphics[width=.235\textwidth]{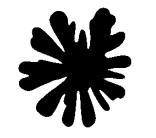}}
  \subfigure[\qquad\qquad\qquad\qquad\qquad\qquad]{\includegraphics[width=.235\textwidth]{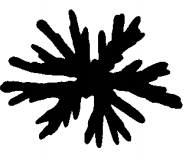}}
  \caption{Examples of (a) Newtonian (water - glycerol) and (b) shear-thinning (water - polyethylene oxide, PEO) Hele-Shaw flow.}\label{fig:exp}
\end{figure}

The classical Hele-Shaw (H-S) experiment~\cite{hele1898flow} is simple: inject a fluid into an immiscible, more viscous fluid confined between two closely-spaced plates. Such a setup is known to be subject to the Saffman-Taylor (S-T) instability~\cite{Saffman}, resulting in complex, fractal-like pattern formation. Extensive reviews of experiments~\cite{maher95}, and of various aspects of theoretical and computational findings~\cite{halsey,casademunt2004viscous,vasil2009hele,homsy87}, are available. Figure~\ref{fig:exp}(a) shows a typical example of experimental results, obtained in NJIT's Capstone Laboratory~\cite{capstone}; many such examples can be found in the literature, see~\cite{maher95} and the references therein.

A key reason for the interest in H-S flow and related instabilities, in addition to the easy visualization, is that its mathematical description is similar to Darcy's formulation modeling porous media flow, which is highly relevant to applications that vary from secondary oil recovery to injection molding~\cite{hieber} and many other natural and man-made setups.  Furthermore, there is a close connection between the S-T problem and the Mullins-Sekerka instability relevant to the propagation of a solidification front in an undercooled liquid~\cite{Langer_rmp80}.  Therefore, there are multiple reasons to work towards understanding the nature of pattern formation and emergent interface morphologies. 

 While a significant body of experimental work has been carried out~\cite{paterson_jfm_1981,vandamme_nature_1986, benjacob_prl_1986, maher_pra_1989,zhao_pra_1992,lindner_jfm_2002} (some with non-Newtonian fluids), it is difficult to reach precise understanding of the morphology of the emerging patterns, for either Newtonian or non-Newtonian fluids. Such morphology is often characterized by the fractal dimension, $D_{\rm f}$, for which values in the range 1.2~-~2.0 have been reported; see, e.g.,~\cite{daccord_prl_1986,zhao_pra_1992,maher95} for discussion. Most of the (relatively) recent experiments seem to converge to a value close to 1.8~\cite{maher95}, but the differences between results are too great to attempt to identify the influence of non-Newtonian effects in the displaced fluid. This influence is crucial to numerous applications; however, due to inherent limitations of the experimental setups, it is difficult to quantify such differences based on experiments alone~\cite{maher_pra_1989,maher95}.

From the theoretical perspective, non-Newtonian free surface flows are nontrivial to model due to often complicated rheological properties. However, the H-S geometry once again comes to the rescue since, in the thin gap limit, for a significant class of non-Newtonian fluids, the most important aspects of their rheology can be reduced to a shear-thinning model involving pressure-gradient dependent viscosity~\cite{KPS96,KSP98,lindner_jfm_2002,benamar_1999}.  Such a simplification was carried out systematically some years back~\cite{FKSP99} showing that, even if elastic effects may be important in general, the thin gap limit reduces the problem to the Darcy-type law ${\bf u} = - b^2\nabla p/[12\mu (|\nabla p|^2)]$, where ${\bf u}$ is the gap (of thickness $b$) averaged fluid velocity, $p$ is the pressure, $\nabla  = (\partial_x,~\partial_y)$, with $(x,y)$ the in-plane coordinates, and $\mu (|\nabla p|^2)$ is the pressure-gradient dependent viscosity.  This simplified viscosity law $\mu (|\nabla p|^2)$ is obtained by gap averaging the shear-rate dependent viscosity $\mu(|{\bf u}_z|^2)$ (here the $z$-subscript represents a partial derivative in the $z$-direction perpendicular to the plates of the cell);  see~\cite{KPS96} for details of how to translate between the two viscosity representations. Such modified Darcy formulations can be obtained by systematic asymptotic expansions applicable to a wide class of non-Newtonian fluids, even when the Weissenberg number quantifying the relevance of elastic effects is $O(1)$.   When combined with the incompressibility condition, $\nabla \cdot {\bf u} =0$, a nonlinear elliptic problem for the fluid pressure is obtained, which is significantly easier to deal with than the original non-Newtonian formulation involving elastic effects. Despite this simplification, it is still challenging and computationally intensive to solve for the complicated interfaces that ultimately develop due to the S-T instability; the problem is of free-boundary type involving in-plane curvature through the Young-Laplace boundary condition $\bar p = - \gamma \kappa$, where $\bar p$ is a suitably renormalized pressure, and $\gamma$ the surface tension; see e.g.~\cite{homsy87} for details. Thus, since carrying out large simulations is difficult, quantifying the emerging pattern morphology and its dependence on non-Newtonian fluid behavior remains elusive. 

An alternative approach is modeling using discrete methods, particularly algorithms based on diffusion-limited aggregation (DLA)~\cite{witten_sander_prb_1983}.  DLA is a version of Monte-Carlo simulation that uses a growing aggregate of particles to simulate solutions of Laplace's equation with free boundaries.  In the context of H-S flow, DLA-type algorithms have been modified to include surface tension effects~\cite{vicsek_prl_1984} and used extensively to discuss the pattern formation that emerges in the S-T instability of displaced Newtonian fluids~\cite{daccord_prl_1986}: it was found that surface tension modifies the fractal dimension $D_{\rm f}$ of the emerging patterns from the ``pure'' DLA value of 1.67 to larger values. Exact values of $D_{\rm f}$ are often unclear, however, and there is a significant body of work discussing the asymptotic value of $D_{\rm f}$ in the limit of large aggregate size, as well as the fractal structure of the emerging patterns~\cite{king_pra_1988,mandelbrot, stanley_pra_1990}. In any case, to the best of our knowledge, whether and to what extent the non-Newtonian character of the displaced fluid influences the value of $D_{\rm f}$, either in experiments or in simulations, is uncertain.

In this paper, we formulate a DLA-type method to model the Hele-Shaw flow of a non-Newtonian fluid. The motivation for this study comes in part from experiments that we briefly report in what follows. The main modeling idea is to incorporate the fluid viscosity's shear rate dependence via a history-dependent pattern growth. We then discuss how such modification influences the morphology of the emerging patterns, focusing for brevity on the fractal dimension of the emerging patterns only.  

The rest of this paper is organized as follows.  Section~\ref{sec:exp} presents examples of experimental results, which serve as a motivation for the development of the models and simulations that follow.  Simulation methods are discussed in Sec.~\ref{sec:methods}, with the implementation details relegated to Appendices~\ref{app:vicsek}~-~\ref{app:fracdimmethods}. The main set of results is presented in Sec.~\ref{sec:results}, with additional discussion related to reproducibility and the influence of modeling parameters in the Appendix~\ref{app:results}.  Section~\ref{sec:conclusions} provides the summary and further discussion. 

\section{Experiments} 
\label{sec:exp}

Figure~\ref{fig:exp} shows two examples of experimental patterns obtained by injecting, in a controllable manner, water into an H-S cell initially containing more viscous fluid, that is either Newtonian (glycerol) (a) or non-Newtonian (polyethylene oxide, PEO)  (b). The H-S cell consists of two plexiglass plates approximately 1/2 inch thick, placed typically at a distance 100-800 $\mu $m apart, with a 2mm wide hole in the top plate.  A syringe is connected to the hole via a plastic tube and water is injected by controlled applied pressure. Both applied pressure and the spacing between the plates were varied in the experiments; the video recordings are available~\cite{capstone}.  
While the experiments were conducted carefully and consistent results were obtained, we note that they are of limited scope and intended only to illustrate visual clues suggesting that the non-Newtonian response of PEO modifies the morphology of the evolving patterns; the interested reader is directed to the extensive review by McCloud \& Maher~\cite{maher95} and numerous references therein showing consistent findings.  We note in particular that the surface tension between the considered fluids varies as well as their rheological properties, leading possibly to modification of the emerging lengthscales.  We have computed the fractal dimensions, $D_{\rm f}$ of the emerging patterns (using methods discussed later in the text), finding systematically lower values for the non-Newtonian fluid.  However, since computing $D_{\rm f}$ involves consideration of different scales that should span orders of magnitude, it is necessary to generate patterns that are much larger than those shown in Fig.~\ref{fig:exp}. For this purpose, we resort to simulations. 

\section{Simulation Methods}
\label{sec:methods}

The DLA algorithm that provides the basis for this work is relatively simple, particularly the on-lattice version we consider. The Brownian motion required by the DLA model is simulated by a random walk along the four possible directions of the grid. One starts with a seed particle at the center of a lattice, generates a particle (`walker') far away, allows the walker to perform a random walk on the lattice until it arrives at a lattice site adjacent to the seed, where it sticks (with unit probability), transforming this particular lattice site from free to occupied (in the algorithm that we use this transformation is permanent; see App.~\ref{app:vicsek} for more details and references to alternative formulations). 

To simulate the S-T instability in the H-S setup, the basic DLA algorithm was modified by Vicsek~\cite{vicsek_prl_1984} via the sticking probability.  The modification promotes sticking in neighborhoods that contain a large number of occupied sites, since such regions, on average, are characterized by a smaller curvature.  Within this model, the sticking probability is defined by 
\begin{equation}
    p_{\rm Newt}  = A \left( {n_\cap \over n_{\rm {total}}}  - n_0\right) +B,
\label{eq:p}
\end{equation}
where $n_\cap $ denotes the number of occupied sites in some neighborhood of linear dimension $l$ (measured in the number of grid points) centered at the considered sticking location, and $n_{\rm total} = l^2$ (Vicsek~\cite{vicsek_prl_1984} used $l = 17$ and $n_0 = (l-1)/(2l)$; see Appendix~\ref{app:vicsek} for further motivation). The original DLA model is recovered by setting $A = 0$ and $B=1$; choosing  $B\in (0,1)$ and increasing the value of $A$ models a stronger surface tension since this encourages walkers to stick in areas of high local particle density, hence, smaller local curvature. 

We wish to simulate both Newtonian and non-Newtonian flows, which requires changes to the algorithm (discussed in some detail in Appendix~\ref{app:vicsek}); here we provide only an outline. In our Newtonian simulations, we improve Vicsek's algorithm, modifying it to avoid the formation of `holes' (unoccupied sites surrounded by occupied ones), and by promoting azimuthal symmetry and minimizing  grid anisotropy.

Building on this improved Newtonian algorithm, we further adapt it to model interface growth for a non-Newtonian (and in particular, shear-thinning) displaced outer fluid. To this end, we need to modify the sticking probability so as to model shear-thinning behavior.  This can be done by realizing that, instead of a shear-rate-dependent viscosity as discussed in the introduction, we may instead think of a viscosity that depends on the velocity magnitude of the displaced fluid, since the two quantities (shear rate and velocity magnitude) are related (fluid velocity is obtained by gap averaging the fluid shear rate).  In the context of DLA-based simulations, we can then model shear thinning behavior by making the sticking probability velocity-magnitude dependent. This is the basic idea underpinning the selected approach. 

In physical experiments, the particular dependence of viscosity on shear-rate may vary widely. Here, we do not attempt to model any particular fluid; instead (with the above discussion in mind) we assume a simple velocity-dependent non-Newtonian correction to the sticking probability of the form
\begin{equation}
p_{\rm nNewt}=CV^{s}.
\label{eq:pNN}
\end{equation}
Here, $C$ and $s$ are two free parameters, chosen in such a way that $p_{\rm nNewt}$ reaches values comparable to $p_{\rm Newt}$ (one could think of these parameters as modeling the shear-thinning properties of the displaced fluid); while $V$ is the velocity magnitude (speed), inversely proportional to the ``age'' of a particle that has since become part of the aggregate (see App.~\ref{app:vicsek} for details). This means that younger (more recently attached) particles are more likely to attract walking particles. The local age of the aggregate has two components: a ``horizontal'' age $T_H$ and a ``vertical'' age $T_V$, which are determined pointwise on each lattice site. To illustrate the algorithm, suppose the aggregate comprises $k-1$ particles.  Depending on how the $k$th particle attaches, its age is determined differently. If the particle $k$ attaches horizontally, then its vertical age is inherited directly from the occupied lattice site responsible for the attachment (neighbor), and the horizontal age is defined by $T_H=k-T_H^{\rm neighbor}$. For a vertical attachment, the two ages are calculated analogously.  The velocity is then given by the vector $\mathbf{v}=\left [ T_H^{-1},T_V^{-1}\right]$, and  the speed by $V=||\mathbf{v}||$.  There are additional considerations if a walker attaches to more than one neighbor at once; see App.~\ref{app:NNrules} for modifications of our algorithm in such cases.  The probability of attachment that we use for non-Newtonian simulations includes Newtonian and shear-thinning effects through the following expression,
\begin{equation}\label{eq:totalprob}
p_{\rm total}=\begin{cases}
0, & \mathrm{if}\ p_{\rm Newt}+p_{\rm nNewt}<0,\\
1, & \mathrm{if}\ p_{\rm Newt}+p_{\rm nNewt}>1,\\
p_{\rm Newt}+p_{\rm nNewt},  & \mathrm{otherwise},
\end{cases}
\end{equation} 
with $p_{\rm nNewt}$ as defined by Eq.~\eqref{eq:pNN}.

\begin{figure}
\begin{centering}
\subfiguretopcaptrue
\subfigure[]{\includegraphics[width=.235\textwidth]{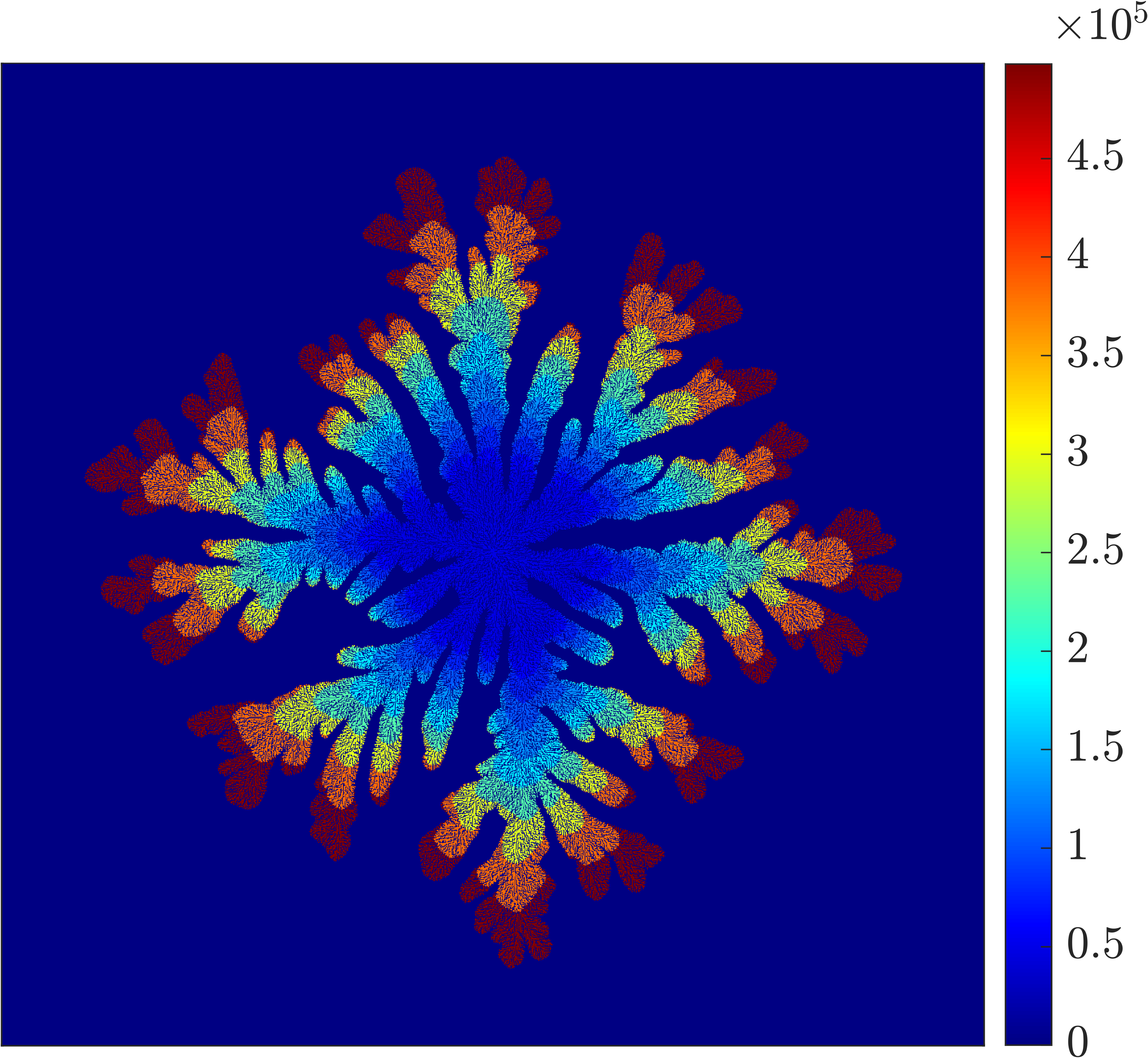}}
\subfigure[]{\includegraphics[width=.235\textwidth]{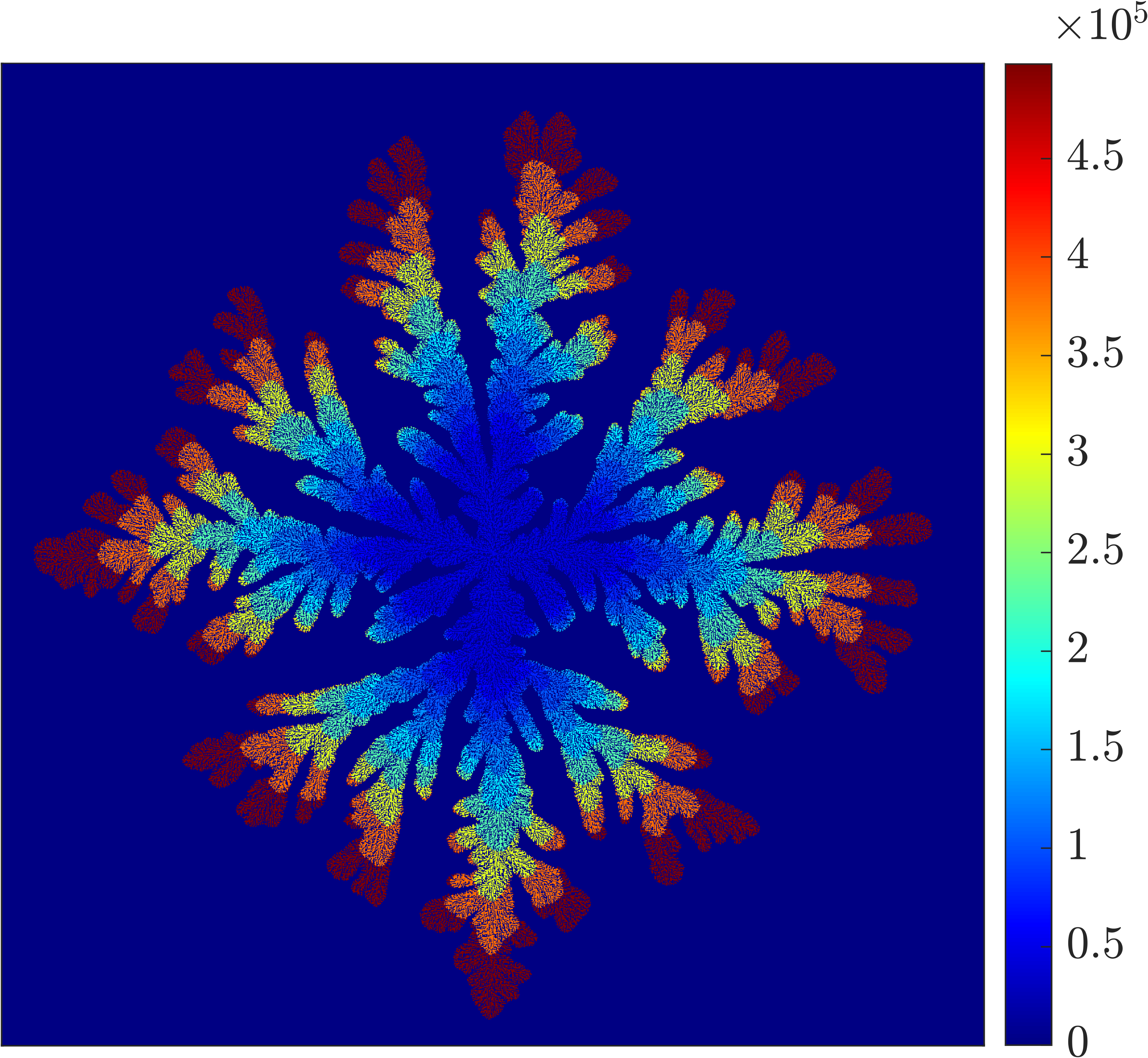}}
\caption{(a) Newtonian and (b) non-Newtonian simulations of pattern growth. Both simulations use $A=4$ and $B=0.4$, while the non-Newtonian simulation also uses $C=0.5$ and $s=0.1$ in Eqs.~(\ref{eq:p},\ref{eq:pNN},\ref{eq:totalprob}). Both plots are on $1800\times1800$ grids; the color palette shows the number of attached walkers. } 
 \label{fig:sims1}
 \end{centering}
\end{figure}

\section{Results}
\label{sec:results}

Figure~\ref{fig:sims1} shows two examples of our Newtonian (a) and shear-thinning (b) simulations.  Visually, these figures suggest the formation of skinnier patterns and possibly different morphology for the shear-thinning case.  Such a trend is consistent with the experimental results shown in Fig.~\ref{fig:exp}, although the differences between the Newtonian and shear-thinning patterns are less pronounced in the simulations.  One reason for this may be the different surface tension between the fluids used in experiments.  Returning to the influence of shear-thinning on pattern characteristics in simulations, the results shown in Fig.~\ref{fig:sims1} suggest that the differences are not simple to quantify, and more importantly, we find that they depend on the pattern size, as observed in previous experiments~\cite{maher95} and simulations~\cite{mandelbrot}.  Our finding, consistent with Newtonian simulations~\cite{mandelbrot}, is that to obtain convergence of the fractal measures quantifying the emerging patterns, millions of walkers are required (the simulation examples in Fig.~\ref{fig:sims1} show $\sim 10^5$ walkers, two-to-three orders of magnitude less than required to obtain converged results for $D_{\rm f}$).  We therefore proceed to discuss the results of much larger simulations and characterize the morphologies of the resulting aggregates by computing $D_{\rm f}$ using box-counting and correlation algorithms; see App.~\ref{app:fracdimmethods} for details. We note that additional measures, such as radius of gyration of the aggregates, were considered but did not provide additional insight and are not reported here for brevity. 
 
\begin{figure}
\centering
\subfiguretopcaptrue
\subfigure[]{\includegraphics[width=0.235\textwidth]{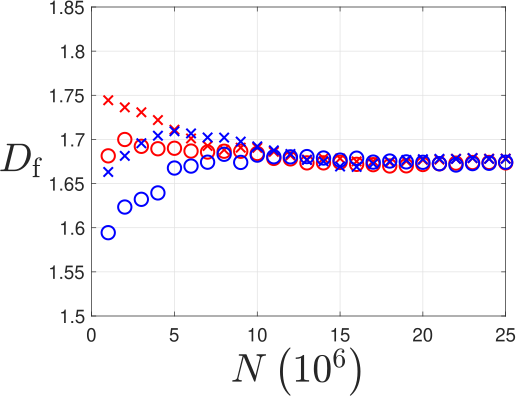}}
\subfigure[]{\includegraphics[width=0.235\textwidth]{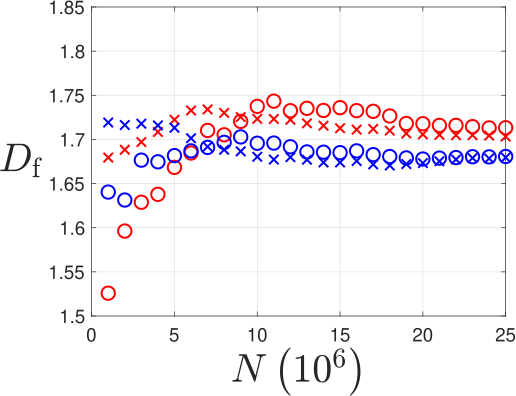}}
\subfigure[]{\includegraphics[width=0.235\textwidth]{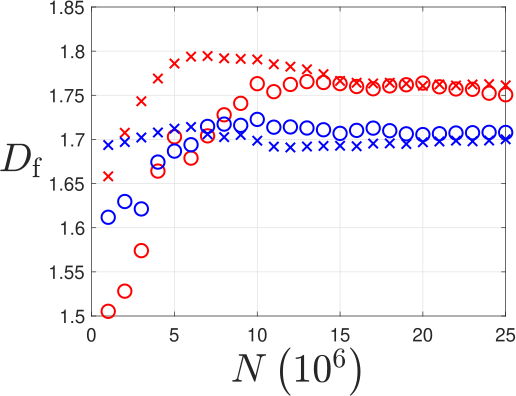}}
\subfigure[]{\includegraphics[width=0.235\textwidth]{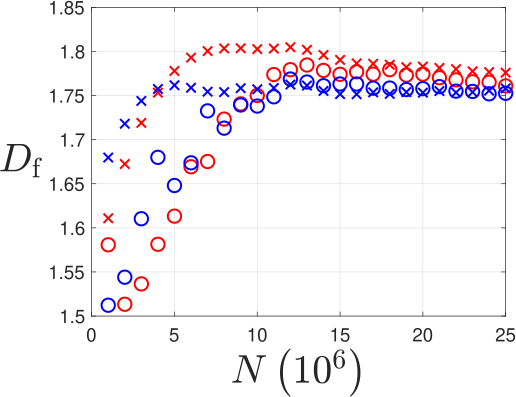}}
\caption{Fractal dimensions using Box Counting (``$\circ$") and Correlation Dimension (``$\times$"), for Newtonian ({\color{red}red}) and non-Newtonian ({\color{blue}blue}) simulations for 1 to 25 million attached walkers, as the surface tension parameter $A$ is varied ($A = 0$ (a),~1 (b),~2 (c),~4 (d)). For $A=0$ (DLA) we use $B=1$, and for $A \ne 0$, $B = 0.4$; also  for the non-Newtonian cases,  $C=0.9,~ s = 0.07$.}
\label{fig:df}
\end{figure}

Figure \ref{fig:df} shows $D_{\rm f}$ as a function of the number of walkers (measured in millions) as the parameters $A,~B$ (entering the definition of $p_{\rm Newt}$ in (\ref{eq:p})) are modified. First, Figure~\ref{fig:df}(a) shows the results for $A = 0$, without surface tension effects. We observe that large simulations are indeed needed to reach both convergence and an agreement between the two calculations; for fewer than $\sim$10 million walkers, the two measures produce differing results. For sufficiently many walkers, we find the value $D_{\rm f}\approx 1.67$, in agreement with values reported in the literature for on-lattice DLA simulations~\cite{halsey,mandelbrot,Bankar}.  We have confirmed that our results are realization-independent; repeating a simulation with different random seed generators produces consistent $D_{\rm f}$ values (for more than 10 million walkers).  Regarding the influence of the non-Newtonian correction, recall that within the DLA algorithm, the probability of sticking is unity, and therefore, the non-Newtonian contribution provided by $p_{\rm nNewt}$ is irrelevant in this case (see~\eqref{eq:totalprob}). Assuming that (Newtonian) DLA represents well the physical S-T problem with vanishing surface tension, the (perhaps not so obvious) conclusion is that $D_{\rm f}$ is not influenced by the non-Newtonian nature of the displaced fluid in this case.

We now consider the effect of changing the surface tension parameter, $A$. Figure \ref{fig:df}(b~-~d) shows the results for $A = 1,~2,~4$.  As $A$ increases, $D_{\rm f}$ increases for both Newtonian and non-Newtonian simulations, but differently. We identify a gap between the corresponding Newtonian and non-Newtonian fractal dimensions, $\Delta D_{\rm f} = D_{\rm f}^{\rm Newt} - D_{\rm f}^{\rm nNewt}$ and find that the size of $\Delta D_{\rm f}$ depends non-monotonously on the value of $A$, with the maximum value observed for $A=2$, see Fig.~\ref{fig:df}(c). For yet larger values of $A$, $\Delta D_{\rm f}$ is found to disappear completely (figures not included for brevity).   

The results of Fig.~\ref{fig:df} lead naturally to the following question: for a given surface tension parameter $A$, is $\Delta D_{\rm f}$ non-negative for all choices of shear-thinning parameters $(C,s)$?  To investigate this question, we carried out additional simulations with several different $(C,s)$ pairs, simulating fluids with different shear-thinning properties.  The results, given in Table III in App.~\ref{app:cs}, suggest that the answer to this question is affirmative: while the size of $\Delta D_{\rm f}$ depends on $(C,s)$ and on $A$, our numerical experiments always find $\Delta D_{\rm f} \ge 0$.

\section{Conclusions}
\label{sec:conclusions}

In this work, we discuss the fractal dimension, $D_{\rm f}$, of the patterns that form due to the Saffman-Taylor instability of Hele-Shaw flow.  We find that the shear thinning property of the more viscous fluid leads to an overall decrease of $D_{\rm f}$, compared to its Newtonian counterpart.  This result is obtained from large-scale simulations of our proposed model, which modifies the well-known approach based on the DLA algorithm for simulating unstable Hele-Shaw two-fluid flow. Such a result would be difficult to obtain either experimentally or via continuous PDE-based simulations since extremely large patterns are needed to produce accurate and converged results.   

The manner in which $D_{\rm f}$ changes due to non-Newtonian behavior is nontrivial.  First, for vanishing surface tension, non-Newtonian behavior is not relevant since our model reduces to classical DLA simulations.  For very large surface tension, the difference in $D_{\rm f}$ between Newtonian and non-Newtonian simulations becomes small, which may be expected in hindsight since the local curvature effects are dominant.  For intermediate surface tension values, however, we find consistently smaller values of $D_{\rm f}$ for non-Newtonian fluids, compared to Newtonian ones. We hope that our findings will encourage new research, both experimental and theoretical, that will make progress in identifying the general connection between fluid rheological properties and the fractal dimension of the emerging patterns.

\begin{acknowledgments}
We hereby acknowledge three cohorts of NJIT undergraduate students (too many to list) who worked on experiments and preliminary versions of the model presented here, with the last iteration leading to the results given in~\cite{capstone}.
\end{acknowledgments}

\appendix

\section{Monte Carlo simulations: modified Vicsek algorithm} 
\label{app:vicsek}

Section~\ref{sec:methods} provides the basic description of the Monte Carlo-based model that we use for simulating aggregate growth. The algorithm we use can simulate both Newtonian and shear-thinning non-Newtonian Hele-Shaw flow. We will first discuss the Newtonian aspects at the core of this algorithm and delay a discussion of the non-Newtonian modifications until Appendices~\ref{app:sim-param} and \ref{app:NNrules}. 

The algorithm is based on an irreversible growth model known as diffusion-limited aggregation (DLA)~\cite{witten_sander_prb_1983}; see also~\cite{halsey} for a concise review. Briefly, DLA is a random walk on a lattice that leads to the formation of an aggregate from a seed particle. To align this with the Hele-Shaw geometry, the seed particle is placed at the origin of a planar rectangular grid. Subsequent particles are initialized one at a time at infinity (from a relatively large distance in the practical implementation) and randomly walk until they come in contact with the seeded aggregate, at which point they stick irreversibly. This process is then repeated until an aggregate of the desired size is formed.  

To simulate the viscous stress forces present in Newtonian flow, Vicsek uses this DLA approach but with a probability of sticking dependent on the local density of the aggregate~\cite{vicsek_prl_1984}. This sticking probability reflects the Laplace-Young boundary condition, relating the jump in pressure to the local curvature of the interface, in the DLA framework. When in the vicinity of the aggregate, the random walker can detect the aggregate's local density by counting the number of occupied lattice sites within a suitably defined neighborhood, taken to be a square of side length $l\in\mathbb{Z}^+$ and odd, centered at the walker's lattice site (the length is measured in units of grid points). 

Vicsek's sticking probability is given by the three-parameter expression
\begin{equation}\label{eq:pVicsek}
    p_{\rm Vicsek}(n_\cap)  = A \left( \frac{n_\cap}{n_{\rm {total}}}  - n_0\right) +B,
\end{equation}
where $n_\cap $ denotes the number of occupied sites in the walker's neighborhood. The three free parameters are $A\in\mathbb{R}^+,\ B\in[0,1],$ and the odd positive integer $l$, which enters via the $l-$dependent parameters $n_{\rm total} = l^2$ and $n_0 = (l-1)/(2l)$. To ensure that this sticking probability is normalized, we restrict $p_{\rm Vicsek}(n_\cap)$ to the interval $[0,1].$ Vicsek showed that this heuristic approach works well in practice to simulate Newtonian Hele-Shaw flow with suitably chosen parameters $(A,B,l)$~\cite{vicsek_prl_1984}. 

We have made two modifications to Vicsek's approach, which are still within the domain of Newtonian flow. First, the aggregate's topology may become multiply-connected, i.e., holes may form. We implement a simple rule to avoid this physically unrealistic (in the context of viscous flow) scenario. We use the following criteria: if a free walker is found adjacent to the aggregate, we check the eight lattice sites that enclose it. If an adjacent site is occupied and its counter-clockwise neighbor site is not, or vice-versa, we call this a flip. If the number of flips is greater than 3 after checking the entire adjacent perimeter of the particle, then a potential hole has formed; the walker does not stick; it moves back to its immediately previous location, and we perform another random walk until the particle is adjacent to the aggregate once again. 

The second modification we make is to use a different metric to define the local neighborhood of the walker. Instead of the rectangular metric used by Vicsek, we use a circular one. In other words, the walker's neighborhood is now a circle instead of a square. This further requires us to reinterpret the parameters $n_0$ and $n_{\rm total}$ defining Eq.~\eqref{eq:pVicsek}. In the original formulation, Vicsek motivates the choice of these parameters by the fact that $n_0 = (l-1)/(2l)$ corresponds to the number of lattice points contained by a rectangle of size $l\times(l-1)/2$ divided by the total number of lattice points $n_{\rm total}=l^2$. In other words, $n_0$ is the number of lattice points that would be occupied if the walker were in contact with a flat interface. As an example, if $l=5$, then the neighborhood is a $5\times5$ square with $n_{\rm total}=25$ and $n_0=4/10=10/25.$

\begin{figure}[t!]
\begin{centering}
\subfigure{\includegraphics[width=.235\textwidth]{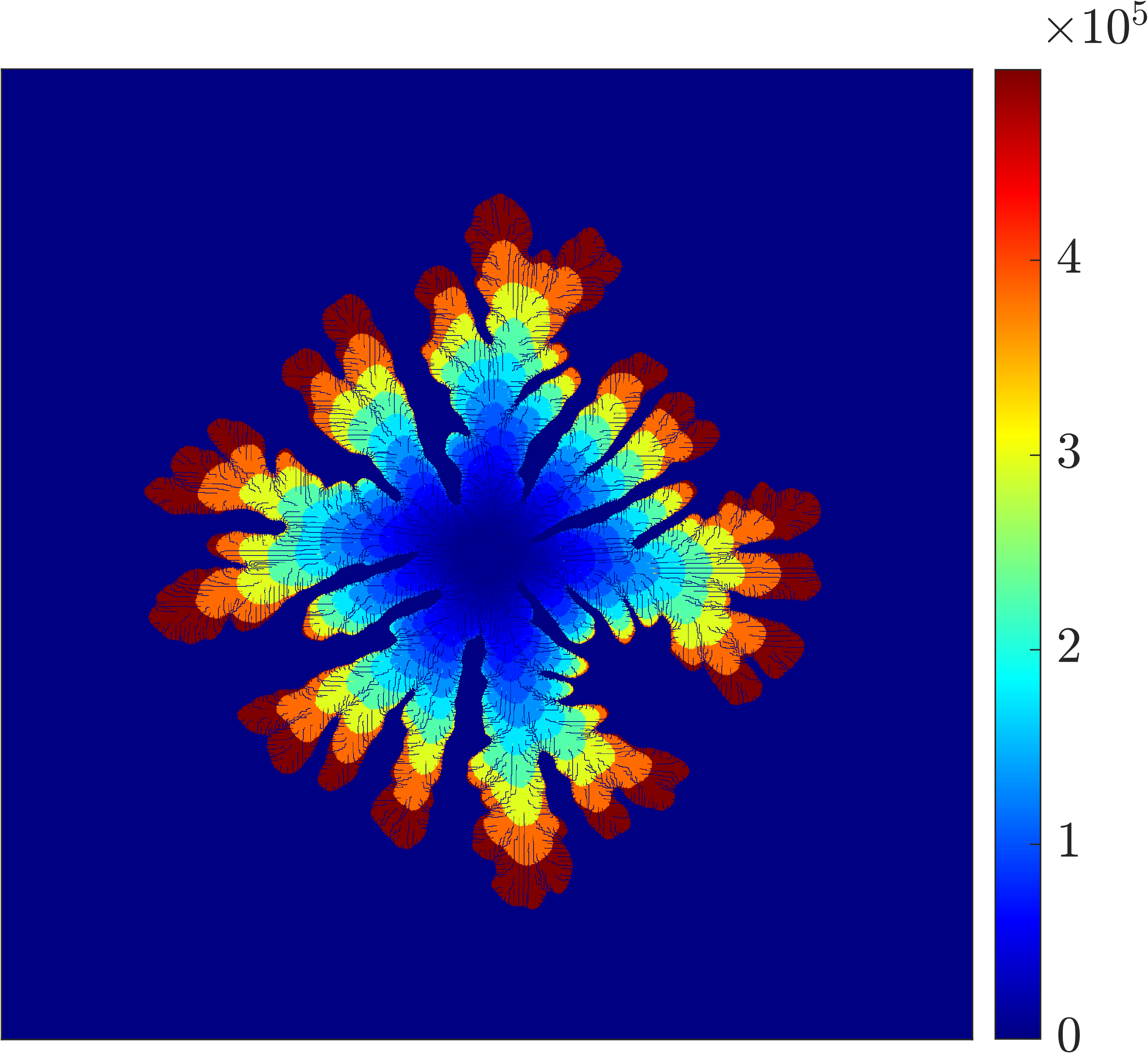}}
\subfigure{\includegraphics[width=.235\textwidth]{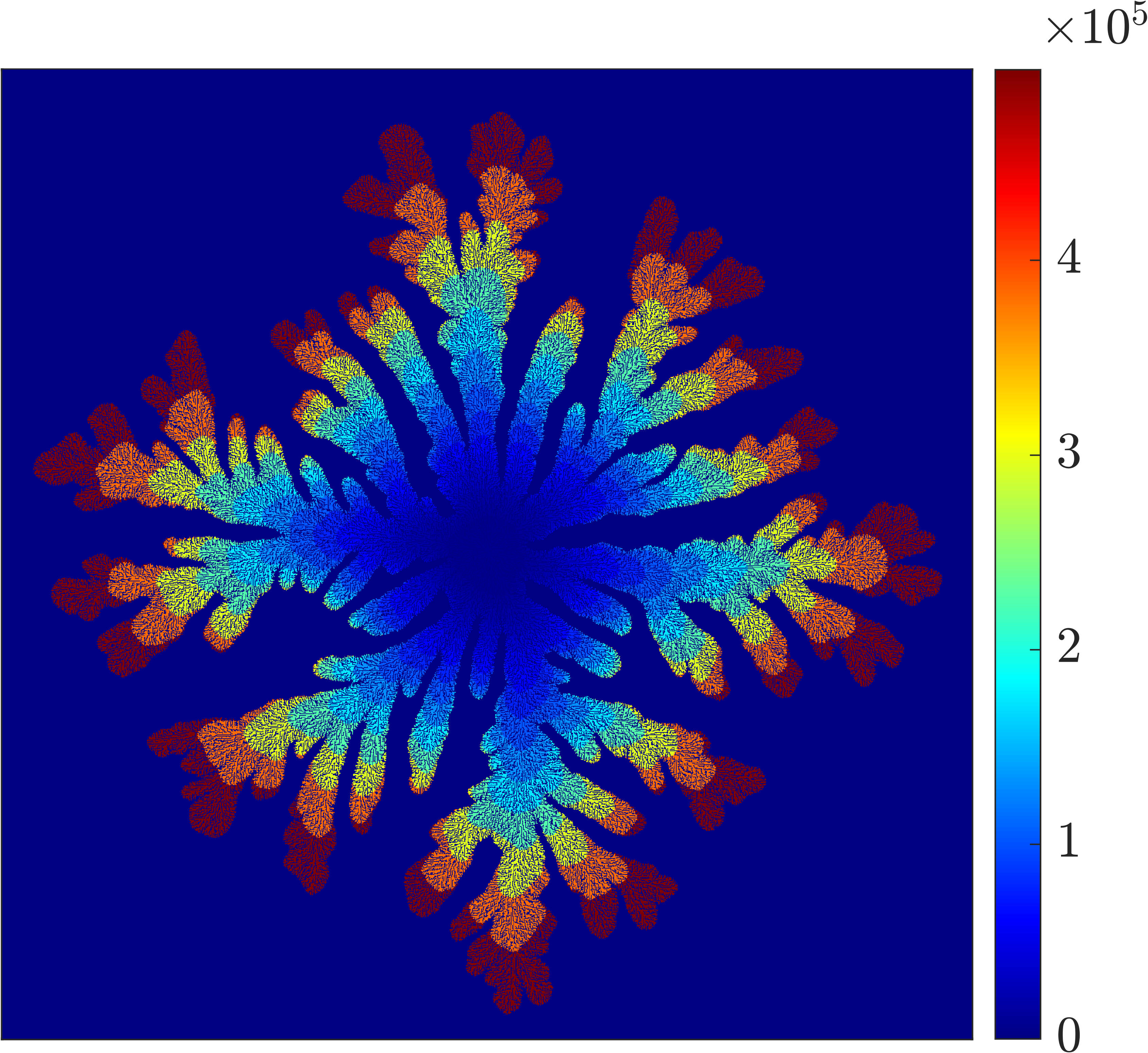}}
\subfigure{\includegraphics[width=.235\textwidth]{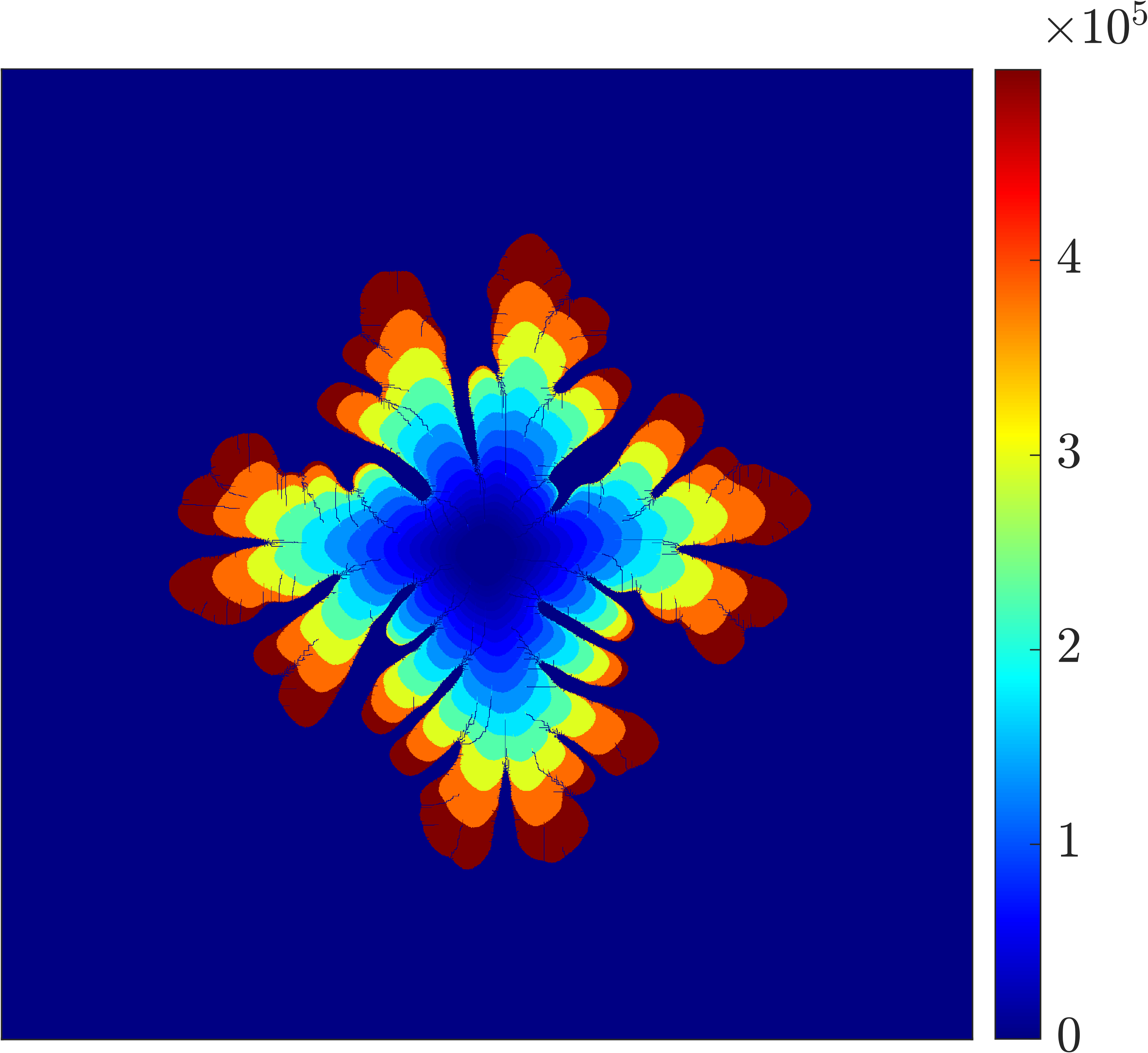}}
\subfigure{\includegraphics[width=.235\textwidth]{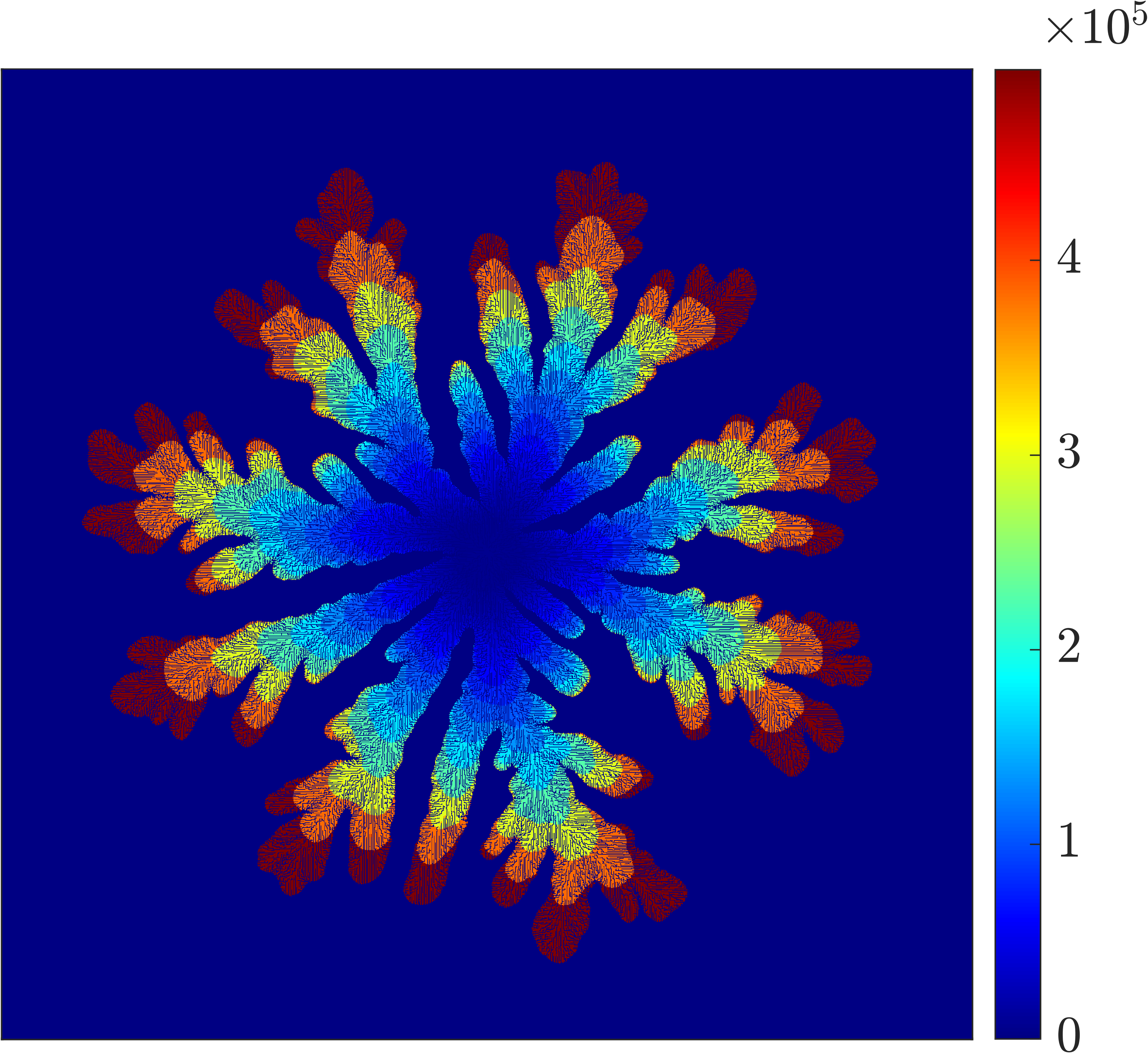}}
\caption{An influence of modifying the neighborhood used in Newtonian simulations. All four simulations have $B=0.4$ and $l=17.$ The top two panels have $A=4$ while the bottom two panels have $A=16.$ The left panels use a square neighborhood, while the right uses a circular one. All plots are on $1600\times1600$ grids.
}
\label{fig:nbddiff}
\end{centering}
\end{figure}

In our modified approach, instead of a rectangle, we use a circle of diameter $l$ centered at the walker position and once again count the number of occupied sites within this circle.  Since this circle is defined on top of the rectangular geometry specified by the grid,  finding a general formula that expresses $n_0$ in terms of $l$ is difficult. For example, for $l=5$, $n_{\rm total}=13$ and $n_0=4/13$. In the case of $l=17,$ a value we use in all simulations throughout our work, we find by direct counting that $n_0=89/{n_{\rm total}}$, with $n_{\rm total}=193$ here and in Eq.~(1). Figure~\ref{fig:nbddiff} illustrates the typical effects of using the modified versus the original algorithm. We find that the modified approaches encourage azimuthal symmetry of the emerging pattern for larger values of the surface tension parameter $A$.

\section{Discussion of simulation parameters}\label{app:sim-param}

\begin{table*}[t!]
  \centering
  \caption{List of parameters used in simulations.\label{key_parameters}} 
  \begin{tabular}{ccc}
  \hline
  Parameter   & Description & (Range of) Value(s) \\ \hline
  $A$ & Surface tension sticking probability     & $\left[1,4\right]$         \\ 
  $B$   & Constant sticking probability     & $[0.4,1]$         \\ 
  $l$   &   Radius of walkers' neighborhood (nbd)   & 17       \\ 
  $n_0$   &  Occupied sites in a semi-circular nbd   & 89/193       \\ 
  $\left(C,s\right)$   & Parameters defining $p_{\rm nNewt}$    & $(0,1)\times(0,1)$       \\
  \end{tabular}
  \end{table*}
  
Equations (2), (3) and the accompanying description explain how we further modify Vicsek's algorithm to account for non-Newtonian effects due to a shear-thinning displaced fluid. We choose the parameters $C$ and $s$ in Eq. (2) in a manner that leads to an appreciable change in the sticking probability. This is done to mimic typical shear-thinning rheology: larger velocity magnitude of the interface leads to a larger sticking probability, modeling smaller viscosity. Different non-Newtonian fluids have different responses to shear, thus, we can think of different values of these parameters as modeling different fluids.  

\begin{figure}
\begin{centering}
\subfigure[]{\includegraphics[width=.235\textwidth]{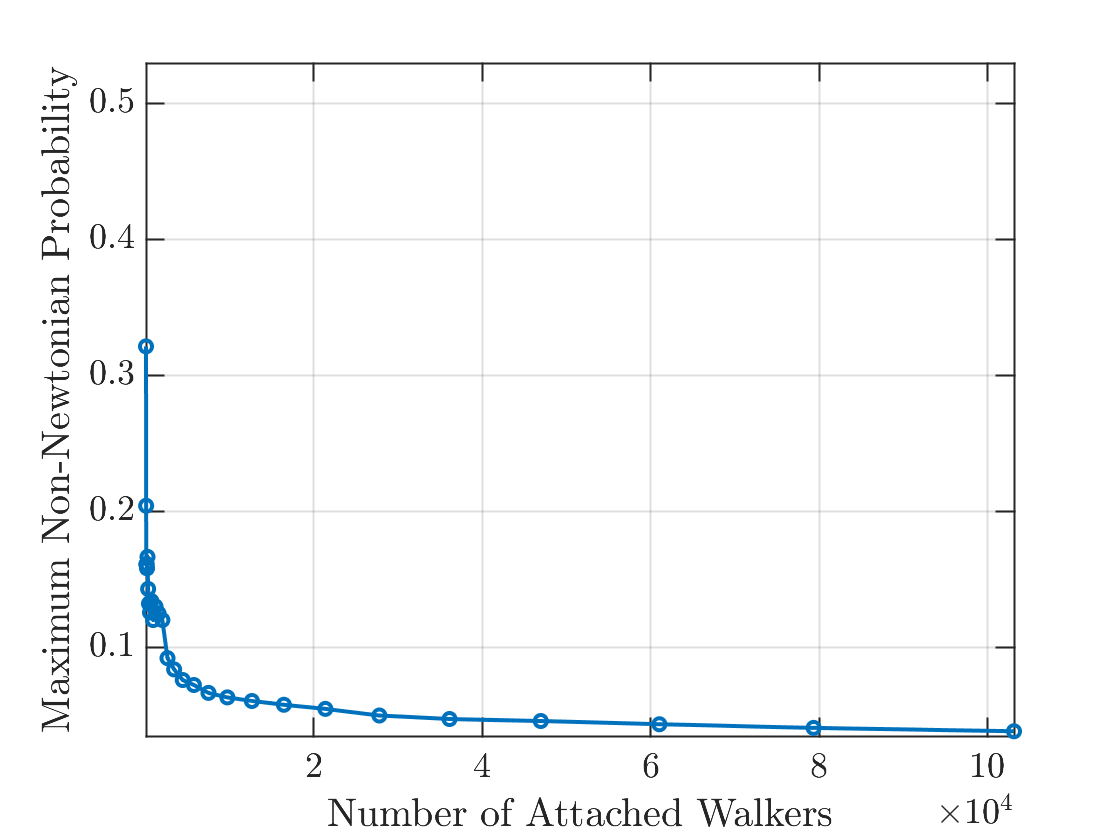}}
\subfigure[]{\includegraphics[width=.235\textwidth]{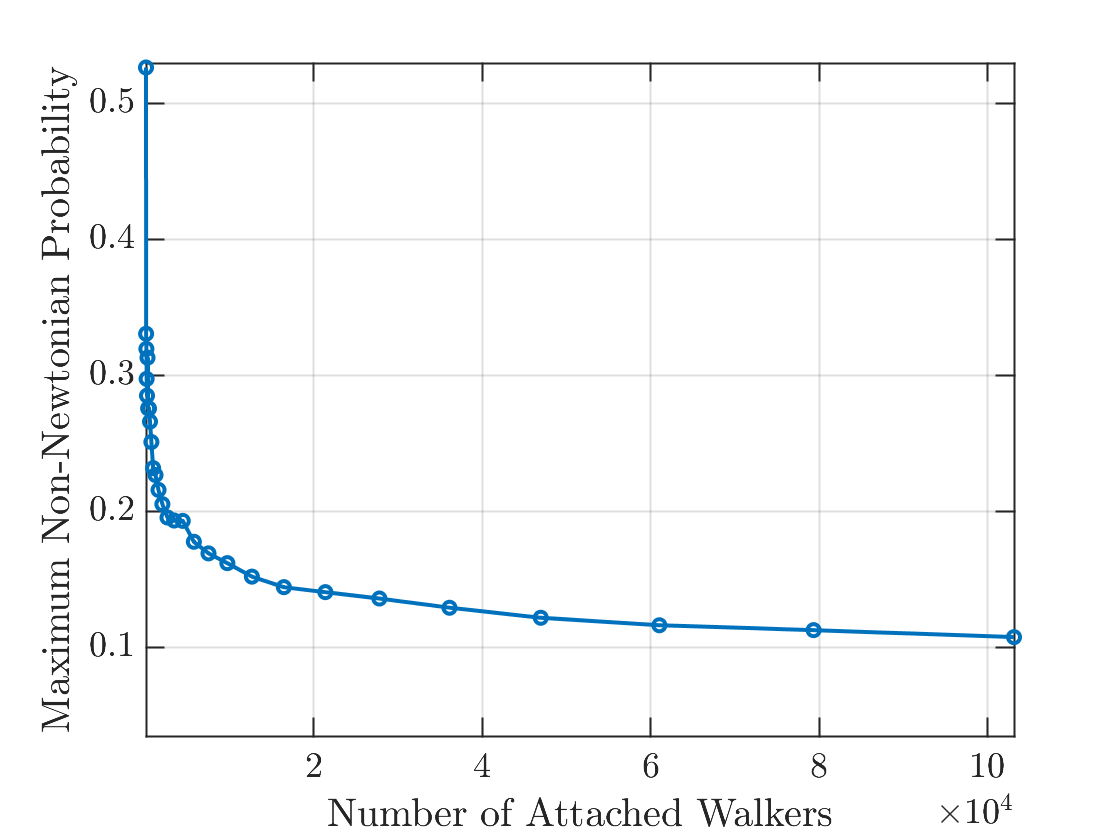}}
\caption{The maximum non-Newtonian probability probability $p_{\rm nNewt}$ of Eq.~(2), running over each period of increasing length shown by the data points. The Newtonian parameters used are $A=4$ and $B=0.4$. Panel (a) uses $C=0.3,$ $s=0.2$ while panel (b) uses $C=0.5,$ $s=0.15$. 
}
\label{fig:corrections}
\end{centering}
\end{figure}

To aid in deciding appropriate parameter values to use in our implementation, we plot the maximum value of the non-Newtonian contribution to the sticking probability, $p_{\rm nNewt}$,  recorded over successively larger periods (labeled by integer $k$) of newly attached walkers (the length of each period is the largest integer of $1.3^{k+1} - 1.3^{k}$, with the data points in Fig.~\ref{fig:corrections} corresponding to $k\in[16,45]$). The choice of $1.3$ and the range of $k$ is arbitrary and was chosen simply for illustrative purposes.

We see that, for the chosen values of parameters $C$ and $s$, the desired qualitative behavior is observed, namely, for small numbers of walkers while the aggregate is increasing rapidly in size, the non-Newtonian correction is large, providing a sizeable non-Newtonian effect. This mimics the large shear-thinning effect anticipated at early times in an experiment, where velocity and shear rate are large. At later times, however, the (now large) aggregate is growing more slowly; one would anticipate a low shear rate in the analogous physical fluid experiment, and the non-Newtonian correction to the sticking probability is accordingly much smaller. 

Table~\ref{key_parameters} summarizes the parameter values used in our simulations.

\section{Further discussion of the attachment rules modeling shear thinning behavior}\label{app:NNrules}

\begin{figure}[htb]
\begin{centering}
\subfigure[]{\includegraphics[height=.235\textwidth]{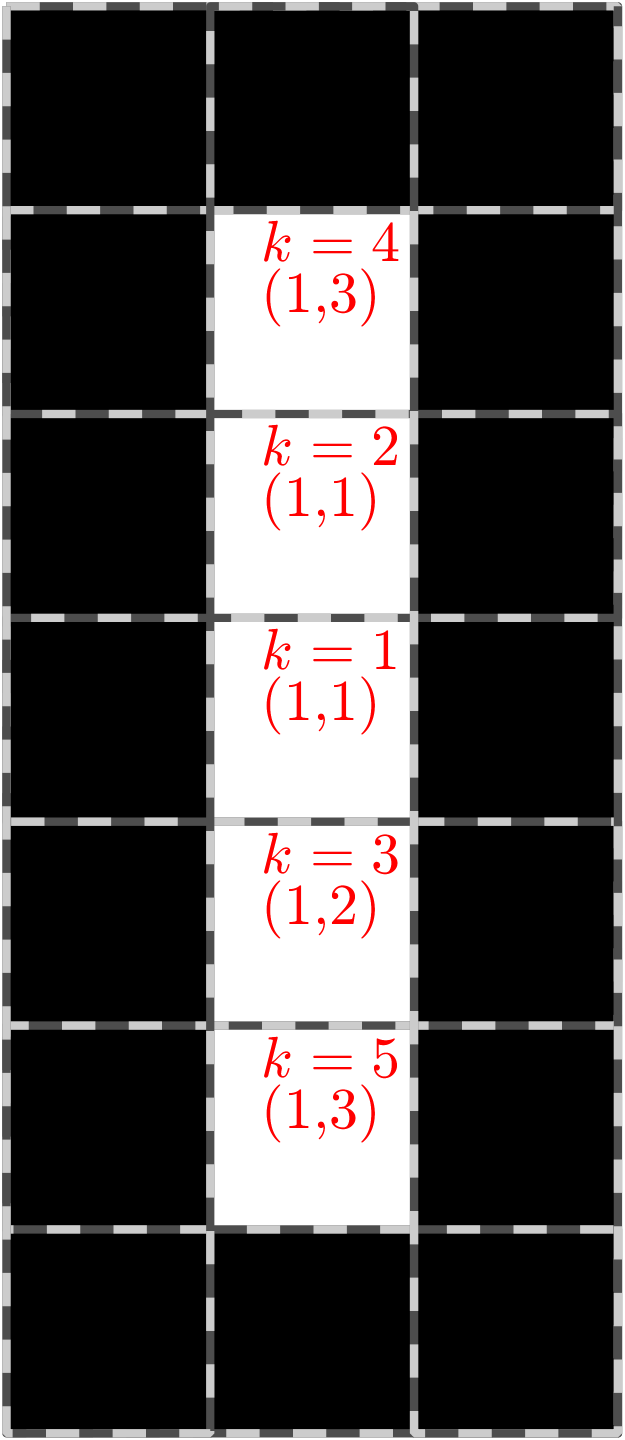}}
\subfigure[]{\includegraphics[width=.235\textwidth]{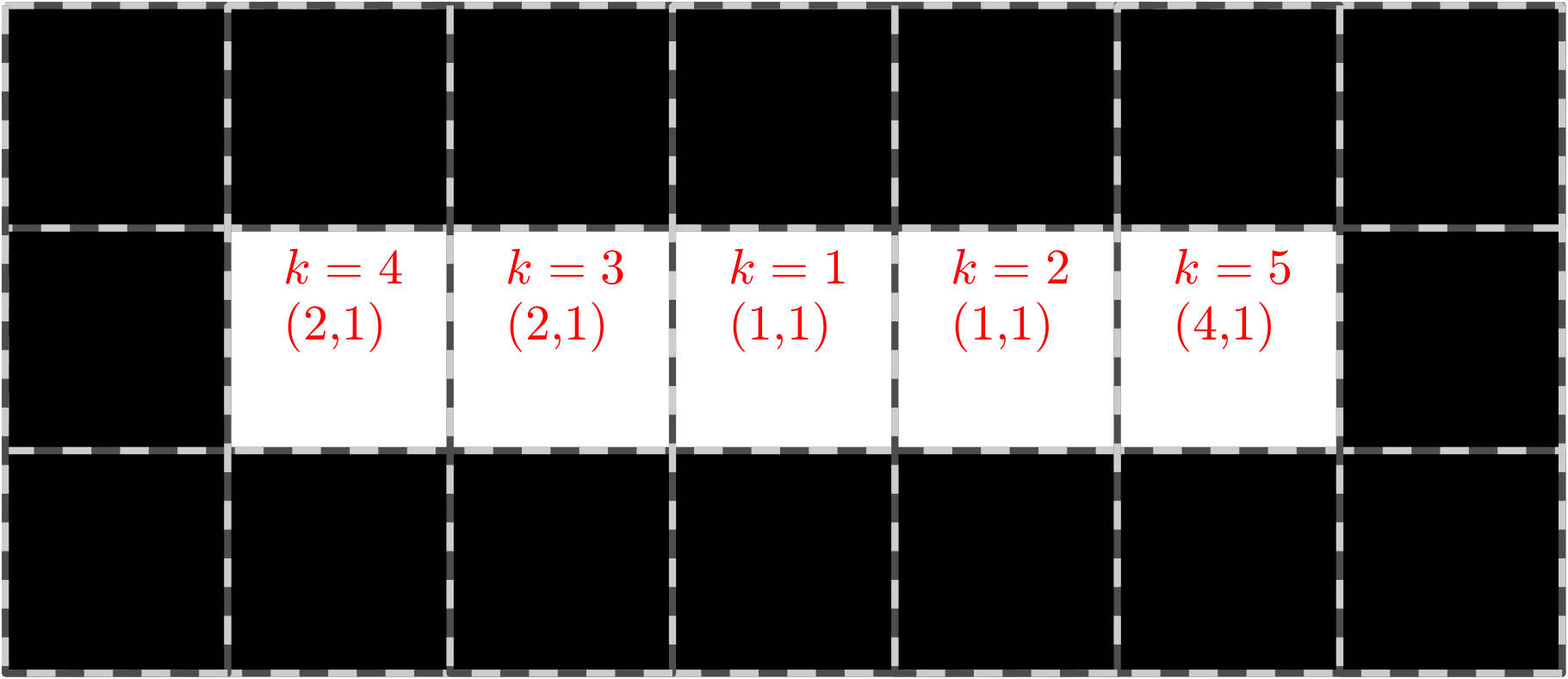}}
\subfigure[]{\includegraphics[width=.235\textwidth]{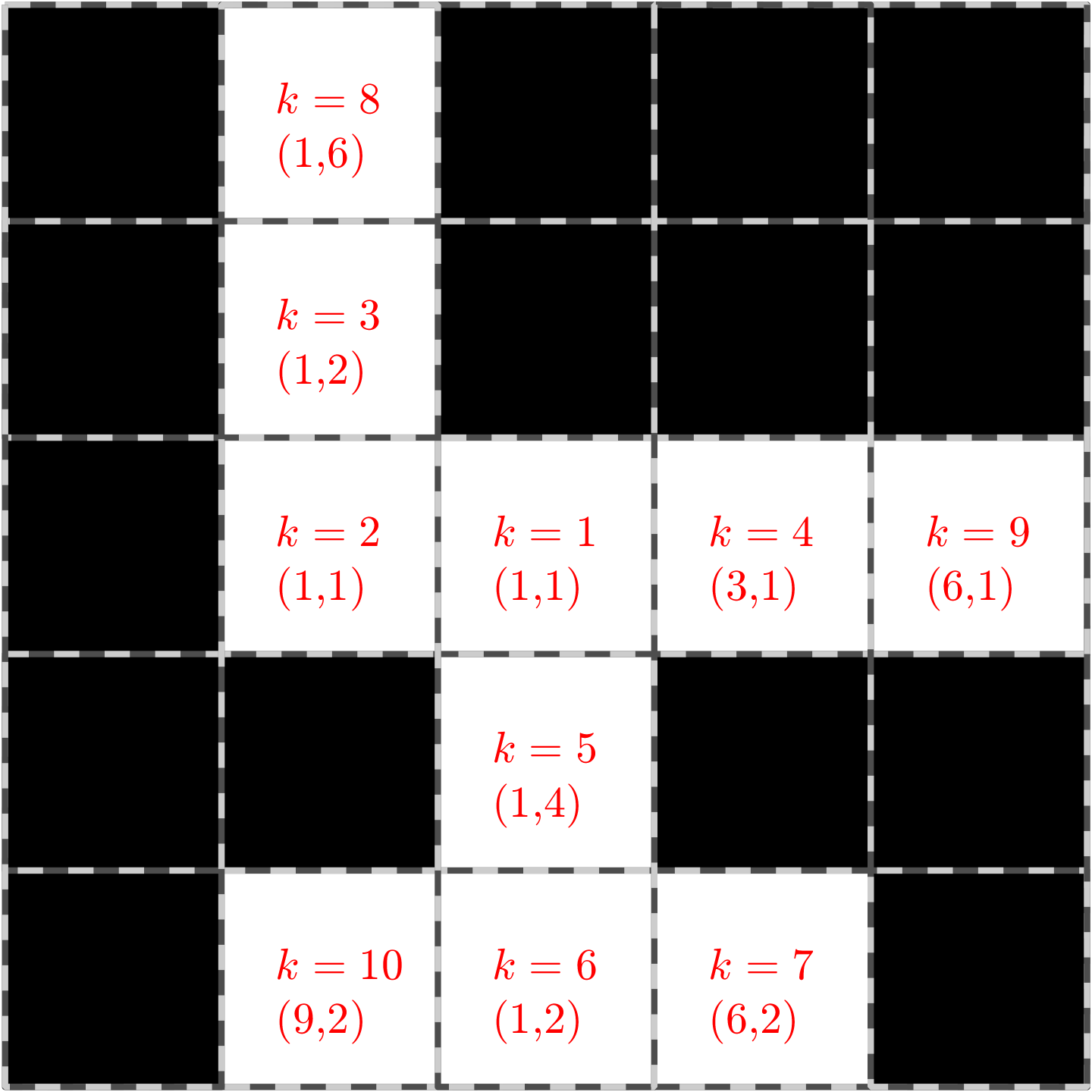}}
\subfigure[]{\includegraphics[width=.235\textwidth]{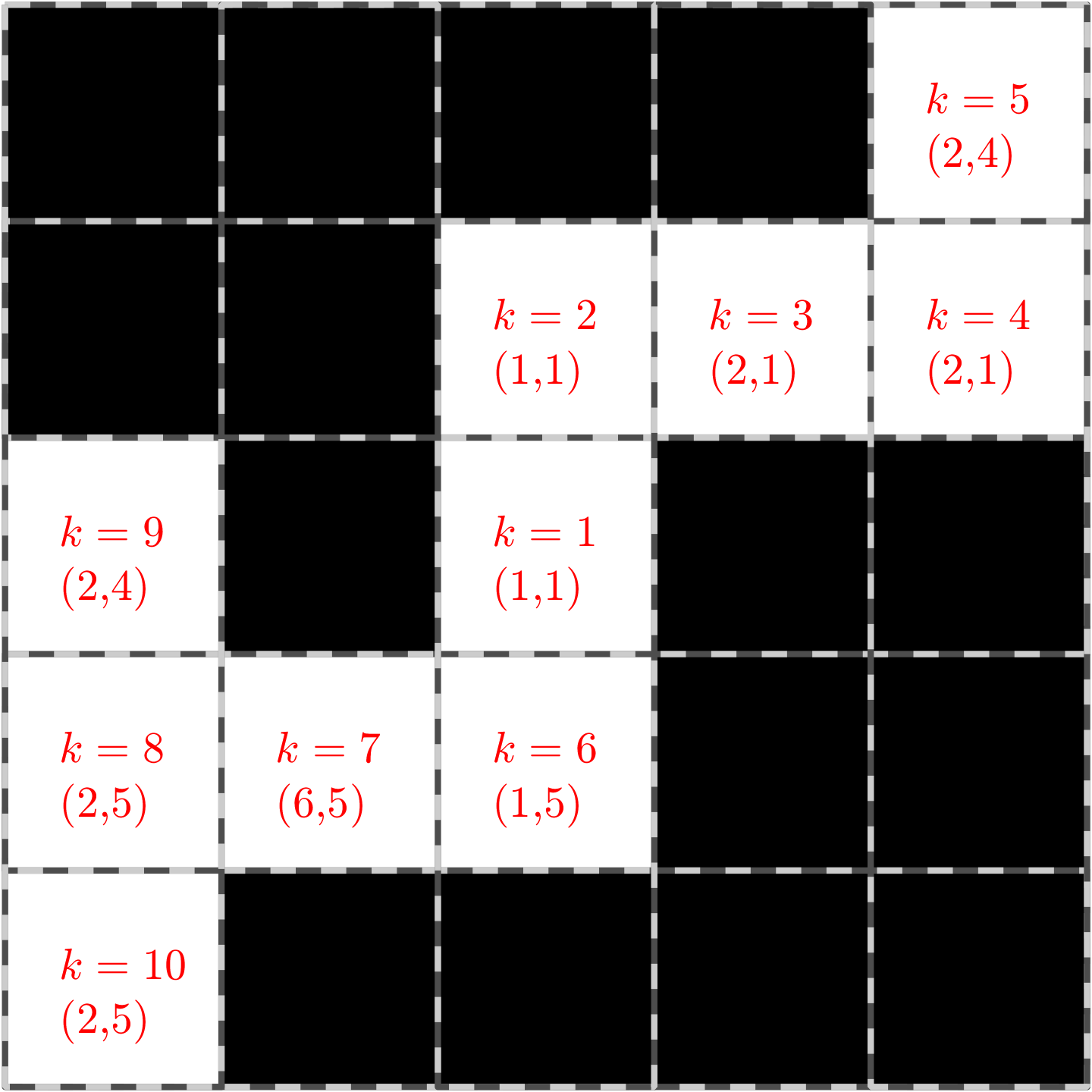}}
\caption{Illustration of attachment rules modeling shear thinning behavior.  White squares represent occupied lattice sites, while black ones represent unoccupied sites. In all panels, age is written in the form $\left(T_H,~T_V\right)$, and the $k=1$ seed particle located at the origin has an age of (1,1). Panels (a) and (b) show simple examples of the horizontal and vertical age evolution of the aggregate in one dimension. Panels (c) and (d) show more elaborate examples where both the horizontal and vertical ages of the aggregate must be accounted for.}
\label{fig:1Dconfig}
\end{centering}
\end{figure}

\begin{figure}
\begin{centering}
{\includegraphics[width=.4\textwidth,]{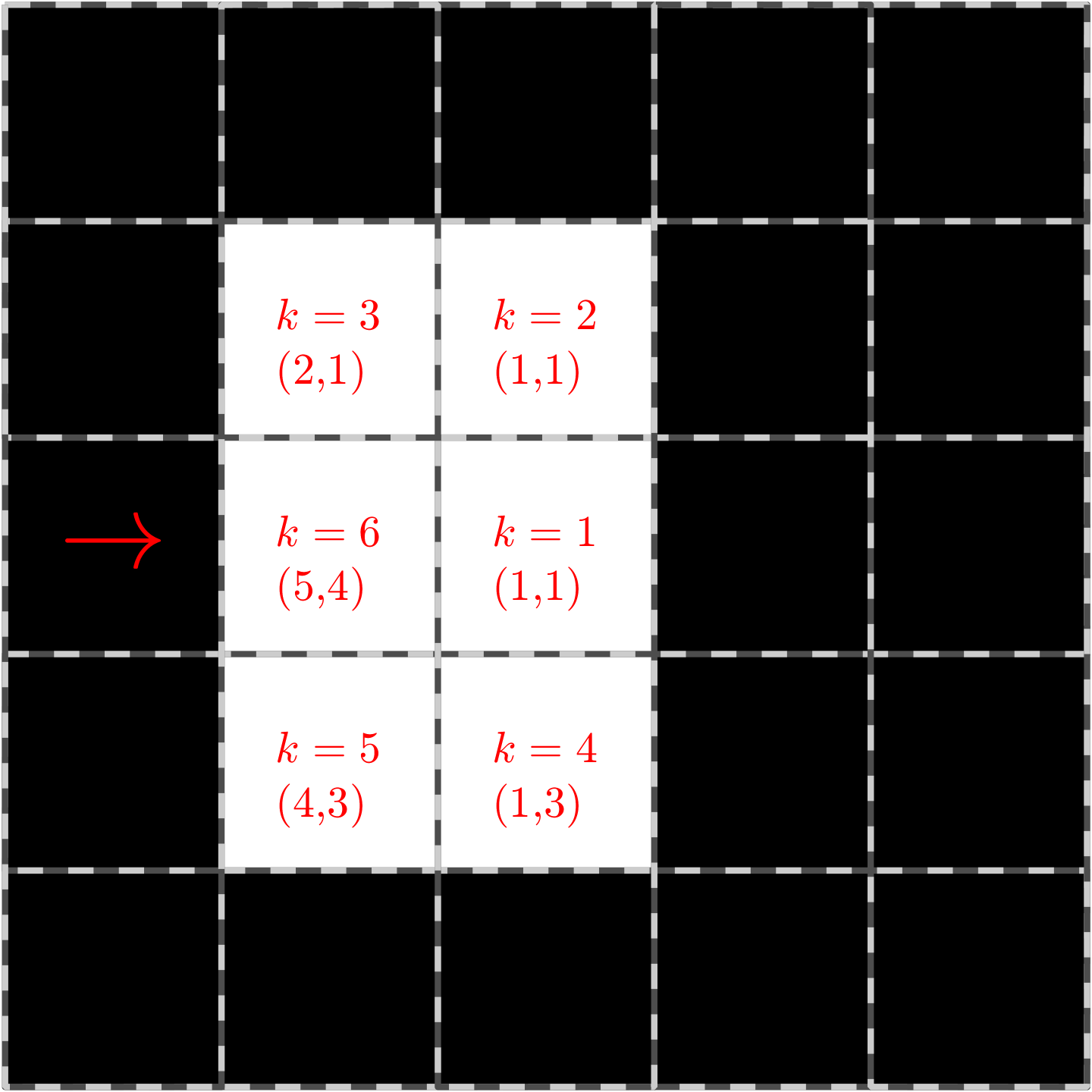}}
\caption{An example where a walker, $(k=6),$ intends to attach at a lattice site in contact with three occupied sites. The red arrow indicates that this walker intends to attach horizontally from the left. In this case, the vertical age $T_V$ is the average of the vertical ages resulting from the $k=3$ and $k=5$ occupied sites, found from $((6-1)+(6-3))/2=4$, while the horizontal age $T_H$, found from $6-1=5$, is updated solely from the $k=1$ site.}
\label{fig:1Dconfig2}
\end{centering}
\end{figure}

In Sec.~\ref{sec:methods}, we discuss the concept of `age'.  Here, we discuss the necessary algorithm modification relevant when the walker is in contact with exactly three occupied lattice sites when it attaches.  Depending on how the walker is attached, the average of either their horizontal or vertical neighbors' ages are used (e.g., for such a horizontal attachment, the horizontal age of the newly-attached walker is inherited from the single neighbor in the aggregate in horizontal contact, while vertical age is obtained by averaging over the two vertical neighbors). Figures~\ref{fig:1Dconfig} and~\ref{fig:1Dconfig2} illustrate the algorithm for a few typical cases.

\section{Calculations of fractal dimension}\label{app:fracdimmethods}

We characterize the morphologies of the resulting aggregates by computing their fractal dimensions. Since the fractal dimension is the crucial measure that we use for the quantification of the emerging patterns, we describe the algorithms implemented in detail here, based on box-counting (Appendix~\ref{app:bc}) and correlation length (Appendix~\ref{app:cl}).

\subsection{Box-counting Dimension}
\label{app:bc}

The box-counting dimension of a bounded set, also known as the Minkowski-Bouilgand dimension \cite{minkowski}, is defined as 
\begin{eqnarray}\label{eq:boxcount}
D_{{\rm f, box}} := \lim_{\epsilon\rightarrow 0} \frac{\log{N\left(\epsilon\right)}}{\log{1/\epsilon}},
\label{box_count_limit}
\end{eqnarray}
where $N\left(\epsilon\right)$ is the number of boxes of side length $\epsilon$ required to cover the bounded set in a metric space. 

In practice, the limit specified by Eq.~\eqref{box_count_limit} is difficult to compute. To approximate it, we consider a nonzero limiting value of $\epsilon$ in Eq.~(\ref{box_count_limit}), and multiply both sides by $\log\left(1/\epsilon\right)$ to obtain a rough estimate
\begin{equation}
    N\left(\epsilon\right)\approx c \epsilon^{-D_{{\rm f, box}}},
\end{equation}
where $c$ is a constant, and $N\left(\epsilon\right)$ can be further understood as the number of $\epsilon$-boxes required to cover the aggregate. Then, we calculate the slope of the log-log linear fit of $N\left(\epsilon\right)$ versus side length $\epsilon$. To facilitate this calculation, we use boxes $B_{\epsilon} (\vec{x})$ centered about a data point $\vec{x}$, defined by the $l^{\infty}$-norm, 
\begin{equation}\label{num_box}
B_{\epsilon}\left(\vec{x}\right)=\left\{ \vec{y}:\left\Vert \vec{x}-\vec{y}\right\Vert_{\infty} \leq\epsilon\right\}.
\end{equation}

The numerical approach is summarized by the Algorithm~\ref{box_alg}. For the domain size $N\times N,$ we use intermediate box sizes ranging from $N/4$ to $3N/4.$
\begin{algorithm}[H]
Computing the Box Counting Dimension
\begin{enumerate}
\item Begin with an on-lattice box $B_{\epsilon}$ of side length $\epsilon$ and with a corner that coincides with the top left corner of the domain.
\item If the box has any particles in it or on the boundary, count it as a cover.
\item Sweep across the domain from left to right, top to bottom, without overlapping. Record the total number of covers $N\left(\epsilon\right)$.
\item Decrease the size of the box $B_{\epsilon}$.
\item Repeat Steps 1-4 until enough data pairs $\left(\epsilon,N\left(\epsilon\right)\right)$ are collected.
\item Perform a linear fit of the logarithm of the number of covers versus the logarithm of the size of the corresponding cover. The magnitude of the best-fit line's slope is the box-counting dimension.
\end{enumerate}
\caption{}
\label{box_alg}
\end{algorithm}
\begin{table*}[t]
\centering
\caption{Key algorithm parameters used in Algorithms~2 and 3.}\label{key_algo_parameters}
\begin{tabular}{cc}
\hline
  Parameter  & Description \\ \hline
  $D_{{\rm f, box}}$ & Box counting fractal dimension         \\ 
  $D_{{\rm f, corr}}$   & Correlation fractal dimension           \\ 
  $\epsilon$   &   Box side length        \\ 
  $N\left(\epsilon\right)$ & Number of boxes of side length $\epsilon$ to cover the aggregate\\ 
  $C\left(\epsilon\right)$   &  The correlation integral       \\ 
  $C_N\left(\epsilon\right)$   & An estimator of the correlation integral       \\
\end{tabular}
\end{table*}

\subsection{Correlation Length-based Dimension}
\label{app:cl}

The second approach we use to calculate the fractal dimension is through the correlation dimension \cite{Strog}. First, we form the correlation integral $C(\epsilon)$ (adopting the nomenclature used by \cite{GP}), understood as the average probability of finding two particles within a ball of radius $\epsilon$, defined via a limit, for $\vec{x}\left(i\right)\in\mathbb{R}^{2}$,
\begin{equation}
C\left(\epsilon\right)=\lim_{N\rightarrow\infty}\frac{1}{N\left(N-1\right)}\sum_{\scriptstyle i,j=1 \atop\scriptstyle i\ne j}^{N}\Theta\left(\epsilon-\left\Vert \vec{x}\left(i\right)-\vec{x}\left(j\right)\right\Vert_{l^2} \right),
\label{eq:corr_int}
\end{equation} 
where $\Theta$ is the Heaviside step function. The correlation dimension $D_{{\rm f, corr}}$ is then defined by 
\begin{equation}
    D_{{\rm f, corr}} = \lim_{N\rightarrow\infty} \lim_{\epsilon\rightarrow 0^{+}} \frac{\log C\left(\epsilon\right)}{\log \epsilon}.
    \label{eq:corr_dim}
\end{equation}
In practice, we estimate $C\left(\epsilon\right)$ by the finite correlation sum, for $\vec{x}\left(i\right)\in\mathbb{R}^{2}$,
\begin{equation}
    C_{N}\left(\epsilon\right)=\frac{1}{N\left(N-1\right)}\sum_{\scriptstyle i,j=1 \atop\scriptstyle i\ne j}^{N}\Theta\left(\epsilon-\left\Vert \vec{x}\left(i\right)-\vec{x}\left(j\right)\right\Vert_{2} \right),
    \label{eq:corr_int_est}
\end{equation}
which is a viable approach due to the large number of particles we simulate, thus producing a very large $N$ and a reasonable approximation of Eq.~(\ref{eq:corr_int}). Therefore, to estimate the correlation dimension using $C_{N}\left(\epsilon\right)$, we use
\begin{equation}
   D_{{\rm f, corr}} \approx \lim_{\epsilon\rightarrow 0^{+}} \frac{\log C_{N}\left(\epsilon\right)}{\log \epsilon}.
   \label{eq:corr_dim_est}
\end{equation}
The approach to computing the correlation sum $C_{N}\left(\epsilon\right)$ is outlined by the following: 
\begin{algorithm}[H]
Computing the Correlation Dimension 
\begin{enumerate}
\item Choose the largest search radius $\epsilon$ from a preset range. \label{al:3.1}
\item Select a random particle $\vec{x}\left(i\right)$. \label{al:3.2}
\item Count the number of unique pairs of particles within a Euclidean $l^2$ distance of $\epsilon$ from the particle at $\vec{x}\left(i\right)$. \label{al:3.3}
\item Remove $\vec{x}\left(i\right)$ from the aggregate (to avoid double counting) and go to step 2 until all particles are used. \label{al:3.4}
\item Compute a total count $C_{N}\left(\epsilon\right)$. \label{al:3.5}
\item Restore aggregate and go to step 1 using a smaller value of $\epsilon$. \label{al:3.6}
\end{enumerate}
\caption{}
\label{al:corr_alg}
\end{algorithm} 

A subtlety in the implementation of the correlation integral calculation is to avoid double counting in the double-indexed sum in Eq. (\ref{eq:corr_int_est});
double-counting arises from the fact that the intersection of $\epsilon$-neighborhoods of $\vec{x}_i$ and $\vec{x}_j$, $B_{\epsilon}\left(\vec{x}_{i}\right)\cap B_{\epsilon}\left(\vec{x}_{j}\right)$ is in general non-empty. To avoid double-counting, after we finish counting for $\vec{x}_{i}$, we remove this particle from the aggregate since we have counted all its unique pairs within its $\epsilon$-neighborhood (step~\ref{al:3.4} of Algorithm~\ref{al:corr_alg}). A technicality of this implementation is that, since this removal procedure alters the aggregate, we must return to the original aggregate every time we change the search radius $\epsilon$.

\section{Additional Results}\label{app:results}

\subsection{Reproducibility}\label{app:reproducibility}

\begin{figure}
\begin{centering}
\subfigure[]{\includegraphics[width=.235\textwidth]
{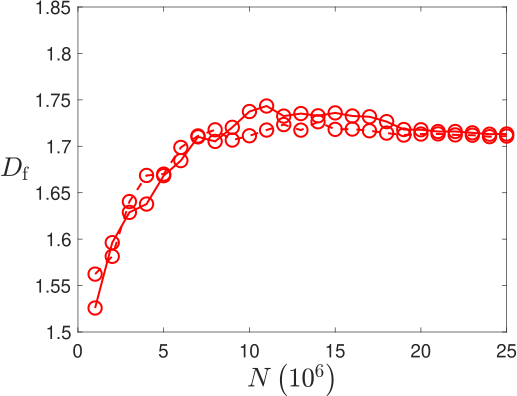}}
\subfigure[]{\includegraphics[width=.235\textwidth]{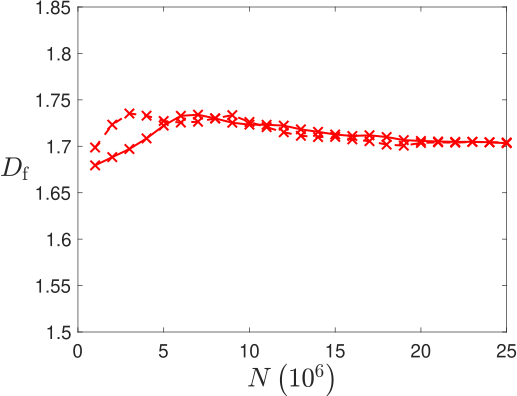}}
\subfigure[]{\includegraphics[width=.235\textwidth]{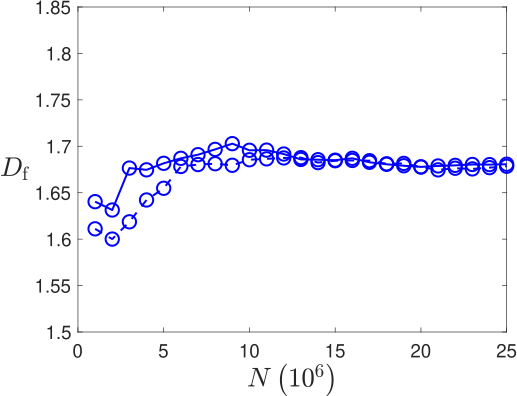}}
\subfigure[]{\includegraphics[width=.235\textwidth]{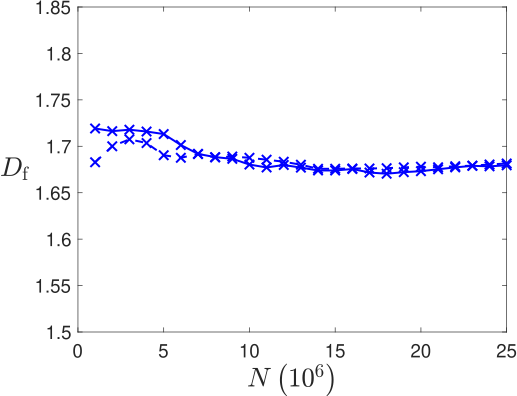}}
\caption{A consistency test for 2 sample simulations with Box Counting (circle ``$\circ$'') and Correlation dimension (``$\times$''). Panels (a) and (b) are Newtonian simulations {\color{red}(red)}. Panels (c) and (d) are non-Newtonian simulations {\color{blue}(blue)} with $C=0.9$ and $s=0.07$. In all cases, we use $A=1$ and $B=0.4$.}
\label{fig:consist}
\end{centering}
\end{figure} 

In this Appendix, we discuss the reproducibility of the fractal dimension results. To show that the results are realization independent, we set the parameters $A=1$ and $B=0.4$, and carry out two independent realizations of Newtonian and non-Newtonian simulations each (non-Newtonian parameters $C=0.9$ and $s=0.07$). Figure~\ref{fig:consist} shows the results of this numerical experiment for both fractal dimension calculation methods. The results, at sufficiently large aggregate sizes, are found to be consistent.

\subsection{Influence of non-Newtonian parameters $(C,s)$} \label{app:cs}

The parameters $(C,s)$ specify shear-thinning response, therefore modeling the rheology of the displaced fluid.  Since we find that $D_{\rm f}$ differs between Newtonian and non-Newtonian simulations, it is not surprising that for different values of $(C,s)$ one could expect to find different values of $D_{\rm f}$.  Table~\ref{fig:cs_table_a1} shows that this is indeed the case; in addition, this table shows that that the change of $D_{\rm f}$, $\Delta D_{\rm f,\dagger} = D_{{\rm f,\dagger}}^{{\rm Newt}} - D_{{\rm f,\dagger}}^{{\rm nNewt}}$ (where $\dagger$ is ${\rm box}$ or ${\rm corr}$) is always non-negative.

\begin{table*}[t]
\centering
\begin{minipage}{.37\linewidth}
    {\bfseries\strut $\left(D_{{\rm f,box}}^{{\rm Newt}},D_{{\rm f,corr}}^{{\rm Newt}}\right)=\left(1.7132,1.7033\right)$}
\begin{tabular}{|c|c|c|c|}
 \hline
$C$    & $s$ & $D_{\rm f, box}^{\rm nNewt}$ & $D_{\rm f, corr}^{\rm nNewt}$ \\
 \toprule
 \hline
0.9000 & 0.07000 & 1.6807 & 1.6795\\ \hline
0.3750 & 0.06250 & 1.6912 & 1.6872\\ \hline
0.2250 & 0.03750 & 1.6902 & 1.6887\\  \hline
0.3750 & 0.1250 & 1.6978 & 1.6932\\ \hline
0.2250 & 0.1000 & 1.7004 & 1.6943\\ \hline
0.2250 & 0.1625 & 1.7038 & 1.7005\\ \hline
0.3000 & 0.1875 & 1.7039 & 1.6995\\ \hline
0.2250 & 0.1750 & 1.7038 & 1.7005\\ \hline
0.3000 & 0.2000 & 1.7030 & 1.7002\\ \hline
0.2250 & 0.1875 & 1.7052 & 1.7026\\ \hline
0.2250 & 0.2000 & 1.7056 & 1.7043\\ \hline
0.3000 & 0.2250 & 1.7058 & 1.7040\\ \hline
0.2250 & 0.2125 & 1.7045 & 1.7030\\ \hline
0.2250 & 0.2250 & 1.7048 & 1.7012\\  \hline
0.3000 & 0.2500 & 1.7060 & 1.7020\\ \hline
\end{tabular}
\end{minipage}%
\hspace*{-0.4in}
\begin{minipage}{.37\linewidth}
    {\bfseries\strut $\left(D_{{\rm f,box}}^{{\rm Newt}},D_{{\rm f,corr}}^{{\rm Newt}}\right)=\left(1.7506,1.7613\right)$}
\begin{tabular}{|c|c|c|c|}
 \hline
$C$    & $s$ & $D_{\rm f, box}^{\rm nNewt}$ & $D_{\rm f, corr}^{\rm nNewt}$    \\
 \toprule
 \hline
0.9000 & 0.07000 & 1.7079 & 1.7002\\ \hline
0.3750 & 0.06250 & 1.6885 & 1.6850\\ \hline
0.2250 & 0.03750 & 1.6917 & 1.6841\\  \hline
0.3750 & 0.1250 & 1.7208 & 1.7159\\ \hline
0.2250 & 0.1000 & 1.7260 & 1.7256\\ \hline
0.2250 & 0.1625 & 1.7349 & 1.7405\\ \hline
0.3000 & 0.1875 & 1.7366 & 1.7419\\ \hline
0.2250 & 0.1750 & 1.7357 & 1.7397\\ \hline
0.3000 & 0.2000 & 1.7376 & 1.7400 \\ \hline
0.2250 & 0.1875 & 1.7342 & 1.7426\\ \hline
0.2250 & 0.2000 & 1.7350 & 1.7406\\ \hline
0.3000 & 0.2250 & 1.7365 & 1.7420\\ \hline
0.2250 & 0.2125 & 1.7370 & 1.7434\\ \hline
0.2250 & 0.2250 & 1.7387 & 1.7435\\  \hline
0.3000 & 0.2500 & 1.7369 & 1.7439\\ \hline
\end{tabular}
\end{minipage}%
\hspace*{-0.4in}
\begin{minipage}{.37\linewidth}
    {\bfseries\strut $\left(D_{{\rm f,box}}^{{\rm Newt}},D_{{\rm f,corr}}^{{\rm Newt}}\right)=\left(1.7609,1.7759\right)$}
\begin{tabular}{|c|c|c|c|}
 \hline
$C$    & $s$ & $D_{\rm f, box}^{\rm nNewt}$ & $D_{\rm f, corr}^{\rm nNewt}$     \\
 \toprule
 \hline
0.9000 & 0.07000 & 1.7526 & 1.7581 \\ \hline
0.5000 & 0.05000 & 1.6833& 1.6792\\ \hline
0.3000 & 0.03000 & 1.6867& 1.6820\\ \hline
0.5000 & 0.1000 & 1.7043& 1.6922\\ \hline
0.3000 & 0.08000 & 1.6960& 1.6883\\ \hline
0.5000 & 0.1500 & 1.7083& 1.6968\\ \hline
0.3000 & 0.1300 & 1.7189& 1.7143\\ \hline
0.3000 & 0.1400 & 1.7263& 1.7293\\ \hline
0.4000 & 0.1600 & 1.7280& 1.7283\\ \hline
0.3000 & 0.1500 & 1.7352& 1.7423\\ \hline
0.4000 & 0.1800 & 1.7407& 1.7479\\ \hline
0.3000 & 0.1600 & 1.7444& 1.7516\\ \hline
0.5000 & 0.2000 & 1.7417& 1.7542\\ \hline
0.3000 & 0.1700 & 1.7456& 1.7561\\ \hline
0.3000 & 0.1800 & 1.7487& 1.7580\\ \hline
\end{tabular}
\end{minipage}%
\caption{Values of $\left(C,s,D_{\rm f, box}^{\rm nNewt},D_{\rm f, corr}^{\rm nNewt}\right)$ for $A=1$ (left), $A=2$ (middle), and for $A=4$ (right). Each table title provides the corresponding box counting and correlation fractal dimension obtained from Newtonian simulations.}\label{fig:cs_table_a1}
\end{table*}

\bibliography{films.bib}
\end{document}